\documentclass{ws-procs9x6}
\usepackage{amssymb}
\usepackage{amsmath}
\usepackage[english]{babel}
\usepackage{graphicx}

\def\Journal#1#2#3#4{{#1} {\bf #2}, #3 (#4)}

\def\RNC{\em Rivista Nuovo Cimento}

\def\PLB{{\em Phys. Lett.}  B}
\def\PRL{\em Phys. Rev. Lett.}
\def\PRD{{\em Phys. Rev.} D}

\def\JETPL{\em JETP Lett.}

\def\JHEP{\em JHEP}
\def\EPHJ{\em Eur.Phys.J}
\def\JPCS{{\em J. Phys.:} Conf. Ser.}

\def\s{{\,\rm s}}

\def\eV{\,{\rm eV}}
\def\keV{\,{\rm keV}}

\def\TeV{\,{\rm TeV}}

\def\({\left(}
\def\){\right)}
\def\cm{{\,\rm cm}}

\def\beq{\begin{equation}}
\def\eeq{\end{equation}}
\def\bea{\begin{eqnarray}}
\def\eea{\end{eqnarray}}

\begin{document}

    \begin{center}
        \large \textbf{DARK ATOMS OF DARK MATTER AND THEIR STABLE CHARGED CONSTITUENTS}
    \end{center}

    \begin{center}
   Maxim Yu. Khlopov$^{1,2,3}$

    \emph{$^{1}$National Research Nuclear University "Moscow Engineering Physics Institute", 115409 Moscow, Russia \\
    $^{2}$ Centre for Cosmoparticle Physics "Cosmion" 115409 Moscow, Russia \\
$^{3}$ APC laboratory 10, rue Alice Domon et L\'eonie Duquet \\75205
Paris Cedex 13, France\\
E-mail: khlopov@apc.univ-paris7.fr}

    \end{center}

\medskip

\begin{abstract}
Direct searches for dark matter lead to serious problems for simple models with stable neutral Weakly Interacting Massive Particles (WIMPs) as candidates for dark matter. A possibility is discussed that new stable quarks and charged leptons exist and are hidden from detection, being bound in neutral dark atoms of composite dark matter. Stable -2 charged particles $O^{--}$ are bound with primordial helium in O-helium (OHe) atoms, being specific nuclear interacting form of composite dark matter. The positive results of DAMA experiments can be explained as annual modulation of radiative capture of O-helium by nuclei. In the framework of this approach test of DAMA results in detectors with other chemical content becomes a nontrivial task, while the experimental search of stable charged particles at LHC or in cosmic rays acquires a meaning of direct test for composite dark matter scenario.

\end{abstract}
\section{Introduction}
It was shown recently
\cite{I,Levels} that new stable
charged particles can exist, if they are hidden in neutral atom-like states.
To avoid anomalous isotopes overproduction, stable particles with
charge $\pm1$ (like tera-electrons \cite{Glashow,Fargion:2005xz}) should
be absent, so that stable negatively charged particles should have
charge -2 only. This possibility cannot take place in SUSY models but a row of alternative models predict such
particles (see Refs. in \cite{Levels}).

In the asymmetric case, corresponding to excess of -2 charge
species, $O^{--}$, they bind in "dark atoms" with primordial $^4He$
as soon as it is formed in the Standard Big Bang Nucleosynthesis. Such dark atoms,
called O-helium ($OHe$), are assumed to be
the dominant form of the modern dark matter, giving rise to a Warmer than
Cold dark matter scenario \cite{Levels,KK2}.

Interaction of OHe with nuclei in
underground detectors can  explain positive results
of dark matter searches in DAMA/NaI (see for review
\cite{Bernabei:2003za}) and DAMA/LIBRA \cite{Bernabei:2008yi}
experiments by annual modulations of radiative capture of O-helium, resolving the controversy
between these results and the results of other experimental groups.

\section{Some features of O-helium Universe}

As soon as primordial helium is formed in the Big bang nucleosynthesis,
 all free $O^{--}$ are trapped by $^4He$ in
O-helium ``atoms" $(^4He^{++} O^{--})$. The radius of Bohr orbit in these ``atoms"
\cite{I,Levels} $R_{o} \sim 1/(Z_{O} Z_{He}\alpha m_{He}) \approx 2
\cdot 10^{-13} \cm $ is nearly equal to the radius of helium nucleus.

Due to nuclear interactions of its helium constituent with nuclei in
the cosmic plasma, the O-helium gas is in thermal equilibrium with
plasma and radiation on the Radiation Dominance (RD) stage, while
the energy and momentum transfer from plasma is effective. The
radiation pressure acting on the plasma is then transferred to
density fluctuations of the O-helium gas and transforms them in
acoustic waves at scales up to the size of the horizon.

At temperature $T < T_{od} \approx 200 S^{2/3}_3\eV$ the energy and
momentum transfer from baryons to O-helium is not effective
\cite{I,Levels} and O-helium gas decouples from plasma. It
starts to dominate in the Universe after $t \sim 10^{12}\s$  at $T
\le T_{RM} \approx 1 \eV$ and O-helium ``atoms" play the main
dynamical role in the development of gravitational instability,
triggering the large scale structure formation. The composite nature
of O-helium determines the specifics of the corresponding warmer than cold dark
matter scenario.

Being decoupled from baryonic matter, the $OHe$ gas does not follow
the formation of baryonic astrophysical objects (stars, planets,
molecular clouds...) and forms dark matter halos of galaxies. It can
be easily seen that O-helium gas is collisionless for its number
density, saturating galactic dark matter. Taking the average density
of baryonic matter one can also find that the Galaxy as a whole is
transparent for O-helium in spite of its nuclear interaction. Only
individual baryonic objects like stars and planets are opaque for
it.

\section{Radiative capture of OHe in the underground detectors}
\subsection{O-helium in the terrestrial matter} The evident
consequence of the O-helium dark matter is its inevitable presence
in the terrestrial matter, which appears opaque to O-helium and
stores all its in-falling flux.

After they fall down terrestrial surface, the in-falling $OHe$
particles are effectively slowed down due to elastic collisions with
matter. Then they drift, sinking down towards the center of the
Earth.
Near the Earth's surface, the O-helium abundance is determined by
the equilibrium between the in-falling and down-drifting fluxes.

At a depth $L$ below the Earth's surface, the drift timescale is
$t_{dr} \sim L/V$, where $V \sim 400 S_3 \cm/\s$ is the drift velocity and $m_o=S_3 \TeV$ is the mass of O-helium. It means that the change of the incoming flux,
caused by the motion of the Earth along its orbit, should lead at
the depth $L \sim 10^5 \cm$ to the corresponding change in the
equilibrium underground concentration of $OHe$ on the timescale
$t_{dr} \approx 2.5 \cdot 10^2 S_3^{-1}\s$.

The equilibrium concentration, which is established in the matter of
underground detectors at this timescale, is given by
\begin{equation}
    n_{oE}=n_{oE}^{(1)}+n_{oE}^{(2)}\cdot sin(\omega (t-t_0))
    \label{noE}
\end{equation}
with $\omega = 2\pi/T$, $T=1yr$ and
$t_0$ the phase.
So, there is a constant concentration and its annual
modulation with amplitude $n_{oE}^{(2)}$.

\subsection{Potential of O-helium interaction with nuclei}

The explanation \cite{Levels} of the results of
DAMA/NaI \cite{Bernabei:2003za} and DAMA/LIBRA
\cite{Bernabei:2008yi} experiments is based on the idea that OHe,
slowed down in the matter of detector, can form a few keV bound
state with nucleus, in which OHe is situated \textbf{beyond} the
nucleus. Therefore the positive result of these experiments is
explained by annual modulation in reaction of radiative capture of OHe
\begin{equation}
A+(^4He^{++}O^{--}) \rightarrow [A(^4He^{++}O^{--})]+\gamma
\label{HeEAZ}
\end{equation}
by nuclei in DAMA detector.

The approach of \cite{Levels} assumes the following
picture: OHe is a neutral atom in the ground state,
perturbed  by Coulomb and nuclear forces of the approaching nucleus.
The sign of OHe polarizability changes with the distance: at larger distances Stark-like effect takes place - nuclear Coulomb force polarizes OHe so that  nucleus is attracted by the induced dipole moment of OHe, while as soon as the perturbation by nuclear force starts to dominate the nucleus polarizes OHe in the opposite way so that He is situated more close to the nucleus, resulting in the repulsive effect of the helium shell of OHe.
When helium is completely merged with the nucleus the interaction is
reduced to the oscillatory potential of $O^{--}$ with
homogeneously charged merged nucleus with the charge $Z+2$.

\begin{figure}
    \begin{center}
        \includegraphics[width=4in]{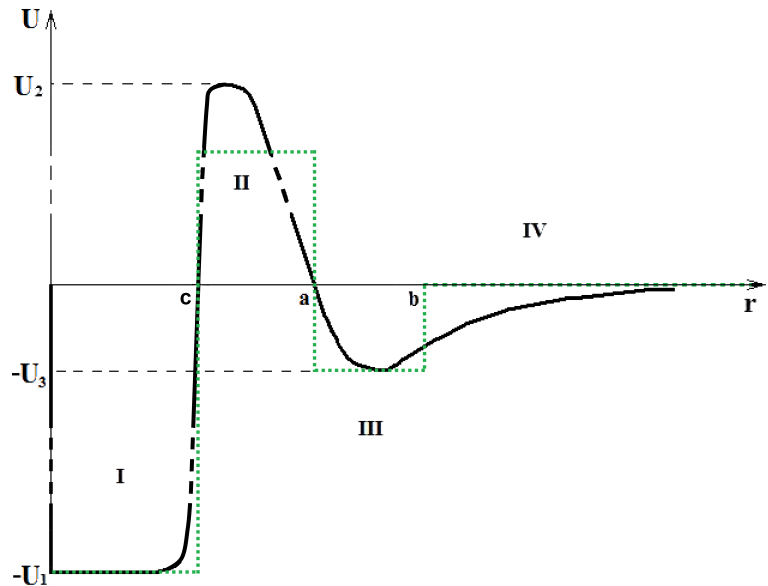}\\
        \caption{The approximation of rectangular well for potential of OHe-nucleus system.}\label{pic1}
    \end{center}
\end{figure}

To simplify the solution of Schrodinger equation the
potential was approximated in \cite{Levels} by a rectangular potential, presented on Fig. \ref{pic1}.

Solution of Schrodinger equation determines the condition, under
which a low-energy  OHe-nucleus bound state appears in the region
III.

\subsection{Radiative capture of O-helium by sodium}

%Owing to the high sensitivity of the result to the nuclear physics
%parameters $d_o$, $\mu$ and $g^2$
% demonstrated in our previous paper \cite{Levels} should be considered as
% very preliminary and can be used only for illustration of the trends in our solution.
The rate of radiative capture of OHe by nuclei can be calculated
\cite{Levels} with the use of the analogy with the radiative
capture of neutron by proton with the account for: i) absence of M1
transition that follows from conservation of orbital momentum and
ii) suppression of E1 transition in the case of OHe. Since OHe is
isoscalar, isovector E1 transition can take place in OHe-nucleus
system only due to effect of isospin nonconservation, which can be
measured by the factor $f = (m_n-m_p)/m_N \approx 1.4 \cdot
10^{-3}$, corresponding to the difference of mass of neutron,$m_n$,
and proton,$m_p$, relative to the mass of nucleon, $m_N$. In the
result the rate of OHe radiative capture by nucleus with atomic
number $A$ and charge $Z$ to the energy level $E$ in the medium with
temperature $T$ is given by
\begin{equation}
    \sigma v=\frac{f \pi \alpha}{m_p^2} \frac{3}{\sqrt{2}} (\frac{Z}{A})^2 \frac{T}{\sqrt{Am_pE}}.
    \label{radcap}
\end{equation}

Formation of OHe-nucleus bound system leads to energy release of its
binding energy, detected as ionization signal.  In the context of
our approach the existence of annual modulations of this signal in
the range 2-6 keV and absence of such effect at energies above 6 keV
means that binding energy of Na-OHe system in DAMA experiment should
not exceed 6 keV, being in the range 2-4 keV. The amplitude of
annual modulation of ionization signal can reproduce the result of DAMA/NaI and DAMA/LIBRA
these experiments for $E_{Na} = 3 \keV$. The
account for energy resolution in DAMA experiments \cite{DAMAlibra}
can explain the observed energy distribution of the signal from
monochromatic photon (with $E_{Na} = 3 \keV$) emitted in OHe
radiative capture.

At the corresponding nuclear parameters there is no binding
of OHe with iodine and thallium \cite{Levels}.

It should be noted that the results of DAMA experiment exhibit also
absence of annual modulations at the energy of MeV-tens MeV. Energy
release in this range should take place, if OHe-nucleus system comes
to the deep level inside the nucleus. This transition implies
tunneling through dipole Coulomb barrier and is suppressed below the
experimental limits.
%Preliminary results give the energy level of
%for $\mu= 320 \MeV$ and $g^2=2$, $\mu= 350 \MeV$ and $g^2=4$, $\mu= 380 \MeV$ and $g^2=10$ or for $\mu= 460 \MeV$ and $g^2=100$.

For the chosen range of nuclear parameters, reproducing the results
of DAMA/NaI and DAMA/LIBRA, our results  \cite{Levels} indicate that
there are no levels in the OHe-nucleus systems for heavy nuclei. In
particular, there are no such levels in Xe, what
seem to prevent direct comparison with DAMA results in
XENON100 experiments. The existence of such level in Ge and the comparison with the results of
CDMS and CoGeNT experiments need special study.
\section{Conclusions}

The results of dark matter search in experiments DAMA/NaI and
DAMA/LIBRA can be explained in the framework of our scenario without
contradiction with the results of other groups.
The proposed explanation is based on the mechanism of low energy
binding of OHe with nuclei. Within the uncertainty of nuclear
physics parameters there exists a range at which OHe binding energy
with sodium is in the interval 2-4 keV. Annual modulation in radiative capture of OHe to
this bound state leads to the corresponding energy release observed
as an ionization signal in DAMA detector.

%The method to calculate the rate of OHe reactions was developed and
%the calculated total amount of such events is shown to be consistent
%with the results of DAMA/NaI and DAMA/LIBRA experiments for the mass
%of OHe around 1 TeV. This method can be applied to the analysis of
%the whole set of inelastic processes, induced by O-helium in matter.

With the account for high sensitivity of the numerical results to
the values of nuclear parameters and for the approximations, made in
the calculations, the presented results can be considered only as an
illustration of the possibility to explain puzzles of dark matter
search in the framework of composite dark matter scenario. An
interesting feature of this explanation is a conclusion that the
ionization signal expected in detectors with the content, different
from NaI, should be dominantly in the energy range beyond 2-6 keV.
Therefore test of results of DAMA/NaI and
DAMA/LIBRA experiments by other experimental groups can become a
very nontrivial task.

The presented approach sheds new light on the physical nature of
dark matter. Specific properties of dark atoms and their
constituents are challenging for the experimental search. The
development of quantitative description of OHe interaction with
matter confronted with the experimental data will provide the
complete test of the composite dark matter model. It challenges search for stable double charged particles at accelerators and cosmic rays as direct experimental probe for charged constituents of dark atoms of dark matter.

%\medskip

\end{document}